\documentclass[prb,twocolumn,showpacs,floatfix,amsbsy]{revtex4}
\usepackage{epsfig,color}
\usepackage[cp1250]{inputenc}
\usepackage{amsmath}

\begin{document}
\title{Singlet-triplet avoided crossings and effective $g$ factor \\ versus spatial orientation of spin-orbit-coupled quantum dots}
\author{M.P. Nowak} \affiliation{Faculty of Physics and Applied
Computer Science, AGH University of Science and Technology, \\
al. Mickiewicza 30, 30-059 Krak\'ow, Poland}
\author{B. Szafran} \affiliation{Faculty of Physics and Applied
Computer Science, AGH University of Science and Technology, \\
al. Mickiewicza 30, 30-059 Krak\'ow, Poland}
\date{\today}

\begin{abstract}
We study avoided crossings opened by spin-orbit interaction in the energy spectra
of one- and two-electron anisotropic quantum dots in
perpendicular magnetic field.
We find that for simultaneously present Rashba and Dresselhaus interactions the width of avoided crossings and the effective $g$ factor
depend on the dot orientation within (001) crystal plane.
The extreme values of these quantities are obtained for [110] and [1$\overline{1}$0] orientations of the dot.
The width of singlet-triplet avoided crossing changes between these two orientations by as much as two orders of magnitude.
The discussed modulation results from orientation-dependent strength of the Zeeman interaction which tends to polarize the spins in the direction of the external magnetic field and thus remove the spin-orbit coupling effects.
\end{abstract}
\pacs{73.21.La}
\maketitle

\section{Introduction}
Spin-related phenomena  in few-electron quantum dots have been under an extensive investigation during the past
decade.  The studies covered spin relaxation involving spin-orbit interaction and phonon emission,\cite{pstano}  spin dephasing \cite{cyw} due to coupling to nuclear spins \cite{petta}  as well as the spin exchange interaction.\cite{sebe} Besides the fundamental interest these studies were motivated by an idea to implement quantum computation \cite{qc} on spins of separate electrons confined in quantum dot arrays. The spin-orbit (SO) coupling is considered for spin manipulation within the orbital degrees of freedom \cite{odf}
as well as in the context of the anisotropy \cite{kavokin} of the exchange interaction for quantum gating.\cite{qg}

Few-electron systems confined in circular quantum dots undergo ground-state angular momentum transitions in external
magnetic field ($B$).\cite{wa,ma,qdr} For the electron pair these transitions are observed only in presence of the electron-electron interaction and are accompanied by spin transitions with ground-state changing between singlet and triplet depending on the parity of angular momentum quantum number.\cite{qdr}
Singlet-triplet ground-state transitions in two-electron quantum dots are observed in charging experiments.\cite{tarucha} In elliptical quantum dots as well as in double dots \cite{harju} the angular momentum transitions are replaced by ground-state parity symmetry transformations still accompanied by singlet-triplet transitions.

The SO coupling mixes the eigenstates of opposite parities and spin. In presence of the SO interaction the singlet-triplet transition occurs
through an avoided crossing that for planar quantum dots was discussed in a number of recent theoretical papers.\cite{anumber}
The SO interaction usually introduces energetically weak effects hence the singlet-triplet avoided crossing is narrow and difficult to observe experimentally. The first observations of singlet-triplet avoided crossings due to SO coupling were performed in transport experiments on quantum dots formed in gated InAs quantum wires.\cite{qws} A transport experiment resolving this avoided crossing in a planar structure was performed only recently \cite{tarucha2} on a single InAs self-assembled quantum dot.

The SO interaction appears due to inversion asymmetry of the crystal lattice (Dresselhaus\cite{dresselhaus} coupling) and/or of the nanostructure (Rashba\cite{rashba} coupling). The resulting SO Hamiltonian is not invariant with respect to rotations within the plane of confinement.
The anisotropy of SO interaction was investigated by observation of singlet-triplet avoided crossing for
rotated external magnetic field vector.\cite{tarucha2}
The study of Ref. [\onlinecite{tarucha2}] extends the previous work \cite{kone} in which the spin splittings were controlled by
orientation of the external magnetic field superposing the effective magnetic field \cite{meier} introduced by SO coupling. It was also demonstrated \cite{rapid} that in presence of the SO coupling the non-linear Kondo conductance depends on the orientation of external magnetic field.

In this paper we consider a planar anisotropic quantum dot in a perpendicular magnetic field and demonstrate that the anisotropy of SO interaction can be
used for tuning the width of the singlet-triplet avoided crossing by spatial orientation of the dot.
This tunability appears provided that both SO coupling types are present.
The discussed effect results from dependence of the effective strength of the Zeeman interaction on the quantum dot orientation within (001) crystal plane.
The Zeeman interaction tends to polarize electron spins in the direction of the magnetic field. A complete polarization amounts in removal of the SO coupling effects.
The extent of the spin polarization -- and thus also the effective Land{\`e} factor\cite{elf} -- vary with the dot orientation.
For similar values of SO coupling constants the width of the avoided crossing changes by two orders of magnitude between a few $\mu$eV to about 0.5 meV. The dependence of the width of singlet-triplet avoided crossing on spatial orientation of the dot is present for any form of confinement potential (quantum well or parabolic profile) for both single and double quantum dots.

\section{Theory}
\subsection{Hamiltonian}
We consider a quantum dot defined within (001) plane with $x$ and $y$ axes identified with [100] and [010] crystal directions,
respectively. The magnetic field is oriented parallel to the growth  [001] direction ($z$).
We adopt a two-dimensional approximation assuming that the confinement potential is
separable into a vertical and planar components $W({\bf r})=V_z(z)+V(x,y)$ and that the vertical
confinement is much stronger than the planar one. Under these conditions the contribution of states excited in the vertical directions
introduced by the spin-orbit-coupling and by the electron-electron interaction is negligible.
The two-dimensional single-electron Hamiltonian takes the form:\cite{anumber}
\begin{eqnarray}
H&=&h\mathbf{1}+\frac{1}{2}g\mu_BB\sigma_z \nonumber \\ & & + H_{SIA} + H_{BIA}, \label{seh}
\end{eqnarray}
where
$h=\left( \frac{\hbar^2 {\bf k}^2} {2m^*} + V(x,y) \right)$
is the spatial Hamiltonian, $\mathbf{1}$ is the identity matrix,
$\mathbf{k}=-i(\frac{\partial}{\partial x},\frac{\partial}{\partial y}) + \frac{e}{\hbar} (A_x,A_y)$ ,
 $g$ stands for the Land{\`e} factor, and $H_{SIA}$ and $H_{BIA}$ describe Rashba\cite{rashba} (
 structure inversion asymmetry) and Dresselhaus\cite{dresselhaus} (bulk inversion asymmetry) SO interactions.
We use the symmetric gauge with $A_x=-y\frac{B}{2}$, $A_y=x\frac{B}{2}$.
 The two-dimensional Rashba interaction
is composed of the diagonal and linear parts $H_{SIA}=H_{SIA}^{lin}+H_{SIA}^{diag}$
with
\begin{equation}
H_{SIA}^{lin} = \alpha(\sigma_x k_y - \sigma_y k_x), \label{lr2d}
\end{equation}
and
\begin{eqnarray}
H_{SIA}^{diag}&=& \alpha_{3D} \sigma_z\left[\frac{\partial W}{\partial y}k_x-\frac{\partial W}{\partial x}k_y\right].
\end{eqnarray}
The two-dimensional coupling constant $\alpha$  in Eq. (\ref{lr2d}) is related to the bulk coupling constant $\alpha_{3D}$
as
 $\alpha=\alpha_{3D}\langle\frac{\partial W}{\partial z}\rangle$, where the average value
 is calculated for the ground-state wave function in the growth direction.
The Dresselhaus interaction contains a linear and cubic terms $H_{BIA}=H_{BIA}^{lin}+H_{BIA}^{cub}$,
    \begin{equation}
H_{BIA}^{lin}=\beta \left[\sigma_x k_x - \sigma_y k_y\right], \label{dr2d}
\end{equation}
 \begin{equation}
  H_{BIA}^{cub} = \gamma_{3D} \left[\sigma_y k_y k^2_x - \sigma_x k_x k^2_y\right],
    \end{equation}
where $\gamma_{3D}$ is the bulk coupling constant and the two-dimensional constant is defined by $\beta= \gamma_{3D} \langle k^2_z\rangle$.
We adopt the material parameters for an In$_{0.5}$Ga$_{0.5}$As quantum dot,
 $\alpha_{3D}=0.572$ nm$^2$ (see Ref. \onlinecite{silva}) and $\gamma_{3D}=32.2$ meVnm$^{3}$ (see Ref. \onlinecite{saikin}),
the electron effective mass $m^*=0.0465m_0$, and the Land\'{e} factor $g=-8.97$.
For $V_z$ in form of a infinite quantum well of height $d$ the two-dimensional Dresselhaus constant is
$\beta=\gamma_{3D}\frac{\pi^2}{d^2}$. For $d=5.42$ nm we have $\beta=10.8$ meV nm. The two-dimensional Rashba constant achieves this value
when (an external or built-in) vertical electric field acquires $188.8$ kV/cm. 

Below we consider a model confinement potential
\begin{eqnarray}
V_c(x',y')& =& -\frac{V_0}{\left( 1+ \left[ \frac{x'^2}{K^2} \right]^\mu \right) \left( 1+ \left[ \frac{y'^2}{L^2} \right]^\mu \right)}, \label{prime}
\end{eqnarray}
where $V_0=50$ meV is assumed for the depth of the quantum dot. The exponent  $\mu=10$ is applied for which the potential profile
has a form of a rectangular potential well with smoothed boundaries. We take $2K=40$ nm as the smaller length of the dot and
the larger length is taken $2L=90$ nm, unless stated otherwise.
The primes standing in Eq. (\ref{prime}) are referred to the crystal directions $x$ and $y$ by
\begin{eqnarray}
x'& = & x\cos(\phi) - y\sin(\phi)\\ \label{obroty}
y' & = & x\sin(\phi) + y\cos(\phi). \nonumber
\end{eqnarray}
The orientation of the dot with respect to the crystal directions is displayed in Fig. 1 for $\phi=\pi/4$.
The effects discussed below remain qualitatively the same for other profiles
of the dots. At the end of next Section we present also results for  elliptical parabolic potential as well as for a double dot potential.
\subsection{Method}
The eigenstates of the single-electron Hamiltonian (\ref{seh}) are calculated in a basis of multicenter Gaussian functions
which is a precise tool for treatment of confinement potentials of arbitrary or no symmetry \cite{chwiej}
\begin{equation}
\phi_\nu=\sum_{ks} c^\nu_{ks} \chi_s \exp\left[-\frac{\left({\bf r}-{\bf R}_k\right)^2}{2a^2}+\frac{ieB}{2\hbar}\left(xY_k-yX_k\right)\right], \label{equ}
\end{equation}
where the centers of Gaussians ${\bf R_k}=(X_k,Y_k)$,
are distributed on a rectangular array,\cite{chwiej} and the localization parameter $a$ is optimized variationally.
In Eq. (\ref{equ}) $s=\pm 1$ and $\chi_s$ are
  eigenstates of the Pauli matrix $\sigma_z$. 

The two-electron states are found by the exact diagonalization approach, which uses
the basis of the anti-symmetrized products of operator (\ref{seh}) eigenstates
\begin{equation}
\Phi=\frac{1}{\sqrt{2}}\sum_{\mu=1}^N \sum_{\nu=\mu+1}^N\left( \phi_\mu (1) \phi_\nu (2) -\phi_\mu (2) \phi_\nu (1)\right).
\end{equation}
for diagonalization
of the two-electron Hamiltonian $H_2=H(1)+H(2)+\frac{e^2}{4\pi\epsilon\epsilon_0 r_{12}}$
($\epsilon=13.55$ is taken for the dielectric constant).
For $2K=40$ nm and $2L=90$ nm we use a basis of $25\times 25 $ centers, which gives
1250 elements including the spin degree of freedom. In the two-electron calculations we take $N=30$ lowest-energy single-electron spin-orbitals which produces
a basis of 435 elements and guarantees the convergence of the variational result better than $1\mu$eV.

\begin{figure}[ht!]
\epsfysize=60mm
                \epsfbox[20 120 590 640] {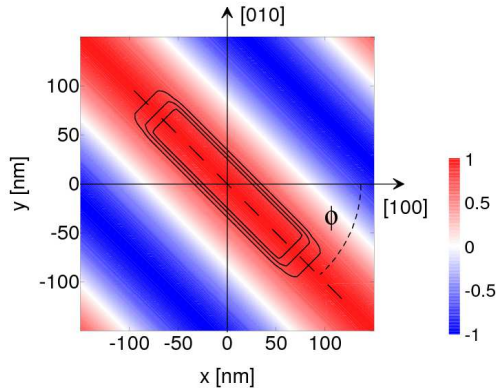}
                 \caption{The contour shows a quantum dot of width $2K=40$ nm and length $2L=200$ nm placed along [1$\overline{1}$0] crystal direction [$\phi=\pi/4$
                 in Eq. (7)]. With the colors we plotted the values of the cosine in the integrand of Eq. (\ref{me2}) for $\alpha=10.8$ meV nm.}
 \label{0eosc}
\end{figure}

\section{Results and Discussion}

\begin{figure}[ht!]
\epsfysize=70mm
                \epsfbox[18 109 579 750] {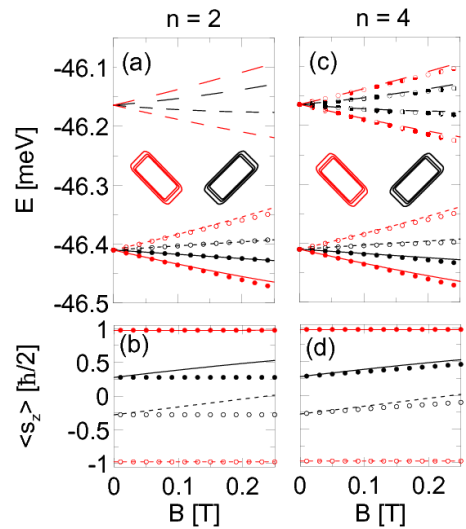}
                 \caption{(a) The dots show the eigenvalues of Hamiltonian (\ref{matrix}) and the lines present the results
                 of diagonalization of the exact Hamiltonian in function of the external magnetic field for $\alpha=\beta=10.8$ meV nm, $2K=40$ nm and $2L=350$ nm.  The results for $[1\overline{1}0]$ ($[110]$) orientation of the dot are displayed in red (black). (b) The spin of the two lowest energy levels for both orientations of the dot. The dots in (c) and (d) show the results of diagonalization of four by four matrix Hamiltonian with basis including the first excited state (see text).}
 \label{1ezeeman}
\end{figure}

\subsection{Effective $g$ factor and orientation of the dot}
In order to explain the dependence of the strength of the Zeeman interaction on the orientation of the dot --
which underlies the results to be presented below --
let us consider the special case of equal linear SO coupling constants $\alpha=\beta=10.8$ meV nm
and neglect the cubic Dresselhaus and diagonal Rashba terms of SO interaction, which are small anyway.\cite{note2}
We consider the approximate Hamiltonian for $B=0$ defined as
$H_0=h{\bf 1} + H_{SIA}^{lin}+H_{BIA}^{lin}$.
 $H_0$ commutes with the operator of [1$\overline{1}$0] spin component and SO coupling shifts
 the entire electron energy spectrum by a constant quantity\cite{com}
$E_N=\epsilon_N-\frac{2\alpha^2m^*}{\hbar^2}$, where $E_N$ and $\epsilon_N$ denote energy eigenvalues with and without SO coupling, respectively.
For $B=0$ the SO coupled eigenfunctions of $H_0$ operator  $\Psi_{N\pm}$ corresponding to $\pm \hbar /2$ spin eigenvalues in the  [1$\overline{1}$0]
direction are related to orbital eigenfunctions $\varphi_N$ that are obtained in the absence of SO coupling as
\begin{equation}
\phi_{N\pm}(x,y)=\frac{1}{\sqrt{2}} \left( \begin{array}{c} 1 \\ \pm e^{-i\pi/4}\end{array}\right) \varphi_N(x,y)e^{\mp\frac{i\sqrt{2}\alpha m}{\hbar^2}(x+y)}. \label{ss}
\end{equation}
The magnetic field vector parallel to the growth direction introduces the Zeeman interaction with $\sigma_z$ matrix to the Hamiltonian, and the $[1\overline{1}0]$ spin component is no longer a good quantum number.
Lets  try to diagonalize the Hamiltonian including the Zeeman effect
$H_z=H_0+\frac{1}{2}g\mu_BB \sigma_z$ taking $H_0$ eigenstates (\ref{ss}) as the basis.
The shortest reasonable basis contains two degenerate ground-state wave functions $\phi_{1\pm}$ corresponding to opposite spin orientations
and the same orbital wave function $\varphi_1$. The  matrix of $H_z$ operator takes the form
\begin{equation}
{\bf H_z}=
\left( {\begin{array}{cc}
 \langle\phi_{1+}|H_z|\phi_{1+}\rangle & \langle\phi_{1 +}|H_z|\phi_{1 -}\rangle  \\
 \langle\phi_{1 -}|H_z|\phi_{1 +}\rangle & \langle\phi_{1-}|H_z|\phi_{1-}\rangle   \label{matrix}
 \end{array} } \right),
 \end{equation}
where both diagonal terms are
$\langle\phi_{1 \pm}|H_B|\phi_{1 \pm}\rangle=E_1 - \frac{2\alpha^2m}{\hbar^2},$
and the off-diagonal ones are
\begin{multline}
\makebox[.55\columnwidth]{$\langle\phi_{1 \pm}|H_B|\phi_{1 \mp}\rangle =$}\\
\makebox[.75\columnwidth]{$\frac{1}{2}g\mu_BB\int|\varphi_1(x,y)|^2e^{\left(\pm i\frac{2\sqrt{2}\alpha m^*}{\hbar^2}(x+y)\right)}dxdy.$} \label{me}
\end{multline}
For potentials with an in-plane inversional symmetry that are considered in this paper the matrix element (\ref{me}) is real and given by
\begin{multline}
\makebox[.55\columnwidth]{$\langle\phi_{1 \pm}|H_B|\phi_{1 \mp}\rangle =$}\\
\makebox[.75\columnwidth]{$\frac{1}{2}g\mu_BB\int|\varphi_1(x,y)|^2\cos\left[{\frac{2\sqrt{2}\alpha m^*}{\hbar^2}(x+y)}\right]dxdy.$} \label{me2}
\end{multline}
Figure 1 shows the plot of the cosine term in the integrand. The argument of the cosine has a fixed orientation with respect to the crystal directions and changes sign along
[110] direction with a period of $\lambda_{SO}=\frac{\pi\hbar^2}{2\alpha m^*}$ (for the applied parameters $\lambda_{SO}=238.4$ nm).
For the quantum dot of length $L=200$ nm oriented along $[1\overline{1}0]$ (see Fig. 1) the cosine has the same sign within the quantum dot area.
For this orientation the off-diagonal terms of Hamiltonian matrix (11) are the largest. On the other hand for the dot oriented along $[110]$ the sign of the integrand oscillates within the quantum dot area and the off-diagonal terms are necessarily smaller. The off-diagonal matrix elements mix the $\sigma_{x-y}$ eigenstates leading to alignment of the spin along the direction of the field ($z$). Therefore, the spin polarization due to the Zeeman effect should be the strongest for $[1\overline{1}0]$ dot orientation and the weakest for the dot oriented along $[110]$ crystal direction.

The results of diagonalization of matrix Hamiltonian (\ref{matrix}) is displayed in Fig. 2(a) for the dot oriented along [110] (black dots) and [1$\overline{1}$0] (red dots) directions. In Fig. 2(b) we displayed the average spin for the two lowest-energy states for both orientations of the dot.
The lines in Fig. 2 show the results of the exact diagonalization with basis given by (\ref{equ}).
The eigenvalues of matrix (\ref{matrix}) quite well reproduce the exact energy levels and the average spin.\cite{note}
According to the intuition given by Fig. 1 the electron spin reacts to the application of the external magnetic field in a more pronounced manner
for [1$\overline{1}$0] orientation of the dot than for the perpendicular orientation [110] . As the non-zero magnetic field lifts the ground-state degeneracy the ground state (the first excited state) becomes nearly spin-up (spin-down) polarized. Polarization of the spin by infinitesimal $B$ for the dot oriented along [110] direction
is much weaker
and increases for higher fields. This increase [black lines in Fig. 2(b)] is not very well reproduced by the two-element basis (black dots).
Inclusion of the first excited  scalar wave function $\varphi_2$ to the approximate calculation gives four basis elements of type (10).
The results are displayed in Figs. 2(c) and 2(d). The four-element basis reproduces also the excited energy levels and an improvement of the description of $s_z(B)$ dependence is obtained particularly for the $[110]$ orientation.

The extent of the spin polarization that varies with the dot orientation results in the
dependence of the Kramers multiplet splitting induced by weak magnetic fields. This in turn leads to orientation dependence of the effective $g^*$ factors\cite{elf} that in the experiments are estimated by the splitting of energy levels by weak magnetic field.
We estimate the effective factor by
\begin{equation}g^*=\lim_{B\rightarrow 0}{\frac{E_2-E_1}{\mu_BB}},\end{equation}
which for the antidiagonal $[1\overline{1}0]$ orientation of the dot gives $g^*=-8.7$ (quite close to $g=-8.97$)
and for the diagonal orientation [110] only $g^*=-2.5$.

\begin{figure*}[ht!]
\epsfysize=65mm
                \epsfbox[18 316 579 542] {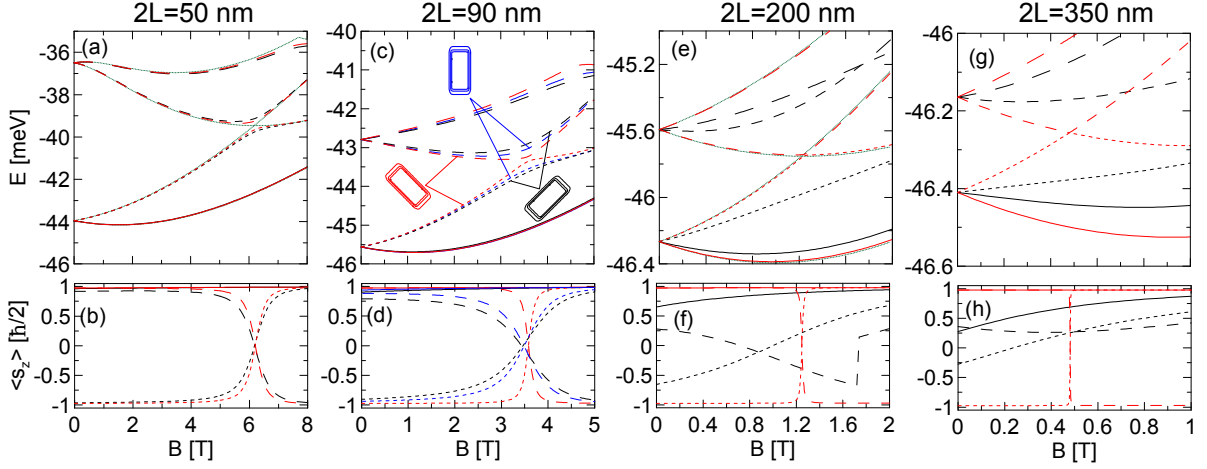}
                 \caption{Energy spectrum (upper row) and mean values of the $z$ component of the spin (lower row of plots) for a single electron
                 in a quantum dot of width $2K$ and various lengths $2L$. The results for the dot oriented along $[110]$ and $[1\overline{1}0]$
                 are given in black and red, respectively. In (c) and (d) we additionally display the results for the dot placed along the [010] direction.  Equal Rashba and Dresselhaus linear coupling constants were assumed $\alpha=\beta=10.8$ meV nm.
                 In (a) and (e) the green dotted lines show the energy spectrum in the absence of SO interaction shifted down on the energy scale by 0.142 meV.} \label{inl}
 \label{1evb0dl}
\end{figure*}

\begin{figure}[ht!]
\epsfysize=70mm
                \epsfbox[26 103 580 700] {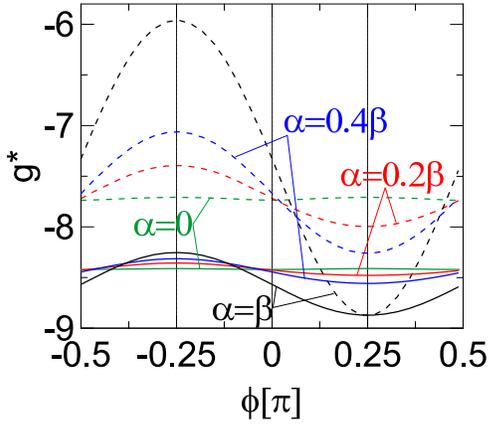}
                 \caption{Effective $g^*$ factor estimated by Eq. (14) in function of the spatial orientation of the dot for $2K=40$ nm, $2L=90$ nm (solid lines)   and $2L=200$ nm  (dashed lines) for $\beta=10.8$ meV nm and various values of $\alpha$.}
 \label{gfactor}
\end{figure}

\begin{figure}[ht!]
\epsfysize=70mm
                \epsfbox[35 125 575 707] {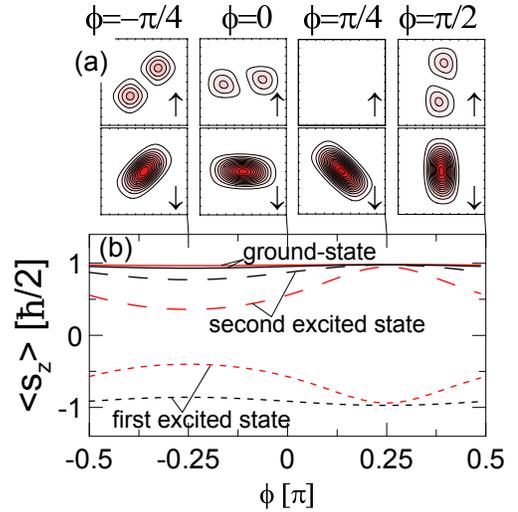}
                 \caption{Results for $2L=90$ nm and $\alpha=\beta=10.8$ meV nm. (a) Spin densities for the first excited-state calculated for subsequent $\phi$ values ($-\pi/4$, $0$, $\pi/4$ and $\pi/2$) for $B=3T$. (b) $\langle s_z\rangle$ for  $B=1$T ($B=3$T) plotted in function of angle $\phi$ with black (red) lines respectively.}
 \label{obrotyr}
\end{figure}

\begin{figure}[ht!]
\epsfysize=90mm
                \epsfbox[73 24 504 822] {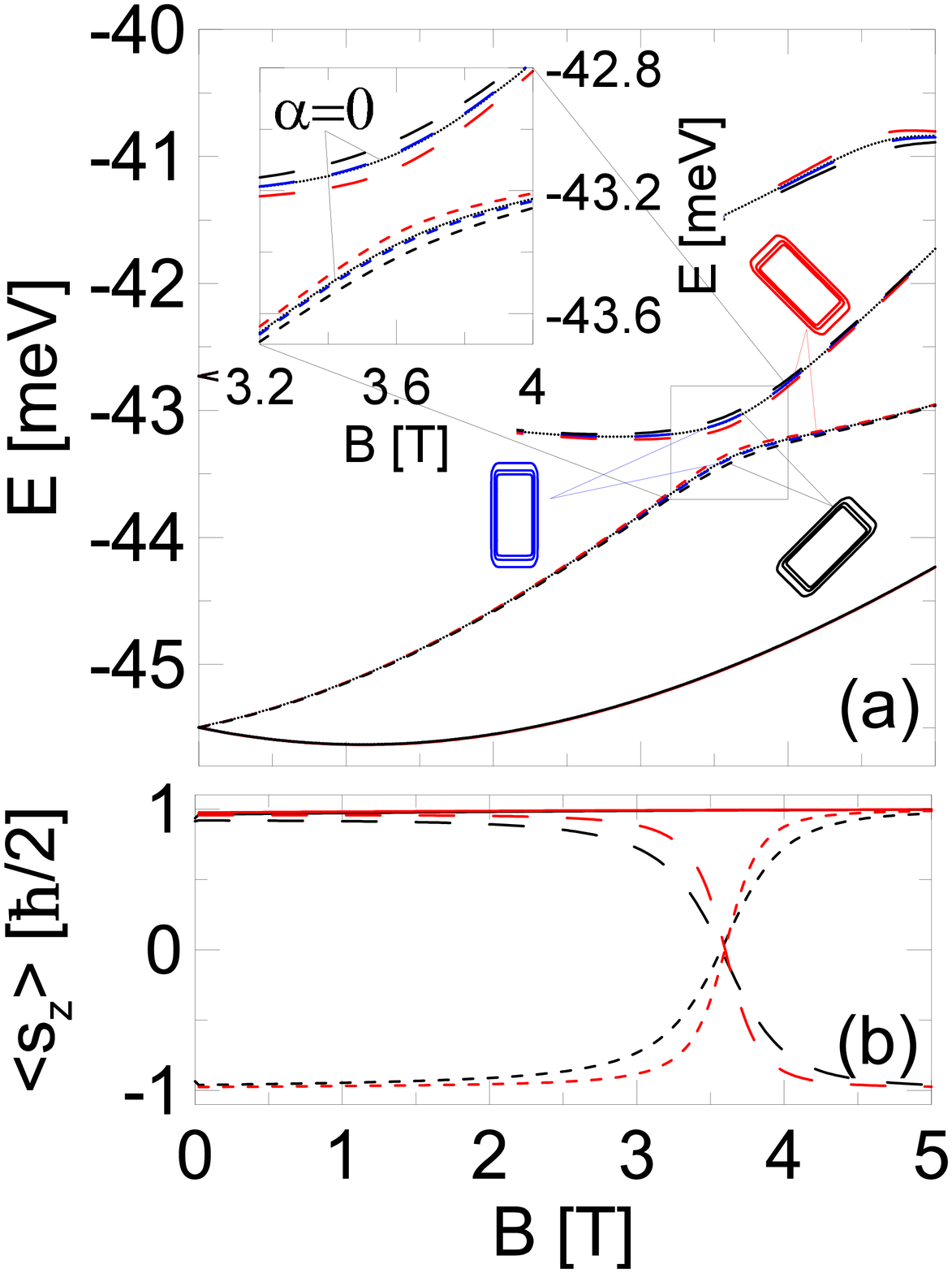}
                 \caption{(a) Energy spectrum for a single-electron dot with a larger length aligned with $[010]$ (blue), $[110]$ (black), and $[1\overline{1}0]$ (red) crystal directions given by solid and dashed lines. The dot size is $2K=40$ nm and $2L=90$ nm, $\beta=10.8$ meV nm and $\alpha=0.2\beta$. The dotted curve shows the results without the linear Rashba term ($\alpha=0$).
                 (b) Mean value of $s_z$ operator for three lowest energy states respectively for [110] and $[1\overline{1}0]$ orientations. The results for this geometrical parameters and $\alpha=\beta$ were given in Fig. \ref{inl}(c,d). }
 \label{02}
\end{figure}

\subsection{Single-electron results}
Let us now consider the results obtained by diagonalization of Hamiltonian (\ref{seh}) with basis (\ref{equ}).
Figure \ref{inl} shows the energy levels and mean spin $z$ components for various length of the dot $2L$ for the diagonal $[110]$ (black color) and
the antidiagonal $[1\overline{1}0]$ (red color) dot orientation.
For the dot which is close to the square profile [$2K=40$ nm and $2L=50$ nm, see Fig. \ref{inl}(a)]
we plotted the results without SO coupling by the green dotted line (shifted down on the energy scale by 0.142 meV).
The first and second energy levels correspond to opposite parity and spin.
For $B$ below the avoided crossing the first excited state is of the even parity with
 spin oriented down  and the second excited state is of the odd parity with spin oriented up.
 The SO coupling opens an avoided crossing between these two energy levels.
This avoided crossing is wider for the diagonal and thinner for the antidiagonal dot orientation  [Fig. \ref{inl}(a)].
The width of this avoided crossing is determined by an extent to which the SO coupling entangles the spin and orbital wave functions.
The width varies more strongly with the dot orientation when the anisotropy of the dot is enhanced,
i.e. for larger length of the dot [Fig. \ref{inl}(a,c,e,g)], particularly when
it becomes comparable to $\lambda_{SO}$. For the antidiagonal orientation of the dots
the strong Zeeman effect quickly polarizes the electron spin and thus removes the SO coupling effects from the energy spectrum.
The energy spectra obtained without SO coupling are close to the ones obtained for the antidiagonal orientation of the dot - see Fig. \ref{inl}(a) and (e).

Figure \ref{gfactor} shows the dependence of the effective $g^*$ factor on the orientation of the dots for $2L=90$ nm  and  200 nm.
The $g^*$ factor acquires maximal (minimal) absolute value for the antidiagonal (diagonal) dot orientation.
For non-equal coupling constants variation of $g^*$ is reduced, and disappears for a single type of SO coupling present.
Note that for the antidiagonal orientation of the dot the same $g^*$ is obtained for both $L$ considered.

For completeness in Fig. \ref{obrotyr}(b) we displayed the mean value of the $z$ component of the spin in function of the dot orientation angle
for $\alpha=\beta$, $2L=90$ nm and two values of the magnetic field $B=1$ T (red) and $B=3$ T (black curves). The spin polarization
is the largest for the $\phi=\pi/4$  and the smallest for $\phi=-\pi/4$.
Fig. \ref{obrotyr}(a) shows the probability density of spin-up and spin-down components for the first excited state.
 Both components of the wave function have opposite spatial parities. For $\phi=\pi/4$ the spin-up component is not visible in the applied contour scale.

The spatial orientation of the dot has a significant influence on the SO related avoided crossings only when both types of the coupling are of
comparable strength. Fig. \ref{02} shows the results for dominant Dresselhaus term $\alpha=0.2\beta$ for $2L=90$ nm. The dependence
of the width of the avoided crossing on orientation is qualitatively the same as for $\alpha=\beta$ [see Fig. \ref{inl}(c,d)], only much weaker.
For comparison the energy spectrum for $\alpha=0$ and $\beta=10.8$ meV nm is given in Fig. \ref{02}(a) with the black dotted line. For a single type of SO coupling present the same energy spectrum is obtained for any dot orientation.

\subsection{Two-electron results}

\begin{figure}[ht!]
\epsfysize=80mm
                \epsfbox[20 100 577 750] {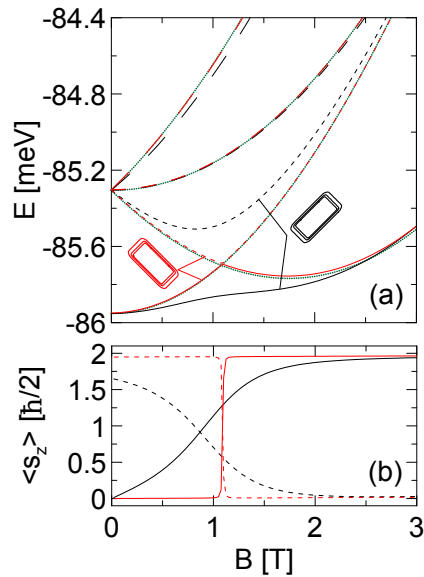}
                 \caption{(a) Two-electron energy spectrum for single-elongated-dot aligned along $[110]$ (black), $[1\overline{1}0]$ (red).
                 Green dotted curve shows the results without SO coupling shifted down by $0.285$ meV. (b) Mean value of $s_z$ for two lowest energy states (solid, short dashed respectively) for [110] (black) and $[1\overline{1}0]$ (red) dot orientation. The value of $\alpha=\beta=10.8$ meV nm is assumed, $2K=40$ nm and $2L=90$ nm.}\label{de}
 \label{2evb0fx01}
\end{figure}

The two-electron spectrum without SO coupling is given in Fig. \ref{de}(a) by the green dotted lines.
When the magnetic field is swept across $B=1.1$ T the ground state changes from the singlet with even spatial parity to the spin-up polarized triplet with odd spatial parity.
The dashed and solid lines in Fig. \ref{de} correspond to SO coupling for $\alpha=\beta=10.8$ meV nm. The results for [110] and $[1\overline{1}0]$ dot orientations are plotted  with the black and the red lines, respectively. For $[1\overline{1}0]$ dot orientation the singlet-triplet transition produces a very narrow avoided
crossing of width 6$\mu$eV as compared to the pronounced (0.37 meV wide) avoided crossing obtained for $[110]$ orientation.
For $[110]$ orientation the mean values of the spin vary smoothly as functions of $B$ [see Fig. \ref{de}(b)], while
for $[1\overline{1}0]$ the mean spin is nearly a bi-valued function of magnetic field [see Fig. \ref{de}(b)], which indicates a removal of the SO coupling effects.
The corresponding energy spectrum [red lines in Fig. \ref{de}(a)] is very close to the spectrum obtained without SO coupling (green dotted lines) up to a constant energy shift [the energy levels without SO coupling are shifted down by 0.285 meV in Fig. \ref{de}(a)].
Similar fact was presented above for the single electron in Fig. 3(a).

\begin{figure}[ht!]
\epsfysize=70mm
                \epsfbox[28 130 577 700] {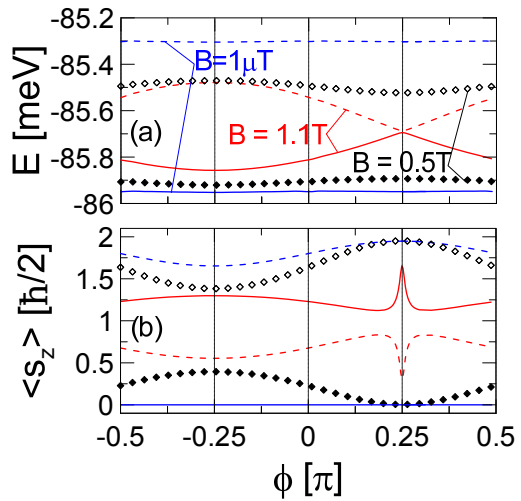}
                 \caption{(a) Two lowest two-electron energy levels
                 plotted in blue for $B=1\mu$T, in black  for $B=0.5$T, and in red  for $B=1.1$T. (b) The $s_z$ mean value. Same parameters as in Fig. \ref{de} were adopted.}\label{2ede}
 \label{2evb0obr.eps}
\end{figure}

In Fig. \ref{2ede}(a) we present two-lowest energy levels as functions of the orientation of the dot for  three values of magnetic field:
1$\mu$T (residual $B$), 0.5 T -- before the singlet-triplet avoided crossing --  and for 1.1 T -- at the center of avoided crossing. The average spins are displayed in Fig. \ref{2ede}(b).
For the residual magnetic field the energy spectrum is independent of the dot orientation and the ground-state spin is zero.
Nevertheless, a dependence of the spin of the  excited state (three-fold degenerate at $B=0$) on the orientation of the dot is noticeable.
For $0.5$ T the energy levels weakly depend on the dot orientation, but dependence of the spins is strong. The situation is opposite for 1.1 T. In both the cases for $\phi=\pi/4$ the spins approach closest $0$ and $\hbar$ -- values that are found in  the absence of SO interaction.

\subsection{Parabolic and double dots}
The profiles of the confinement potentials depend strongly on the type of quantum dots, their size and growth conditions.
The effects discussed above occur in the low-energy part of the spectrum and appear as functions of the dot orientation.
In order to demonstrate that they are not specific to any profile of confinement potential
we considered also a parabolic quantum dot with an elliptical shape and a double dot.

\begin{figure}[ht!]
\epsfysize=40mm
                \epsfbox[17 304 574 570] {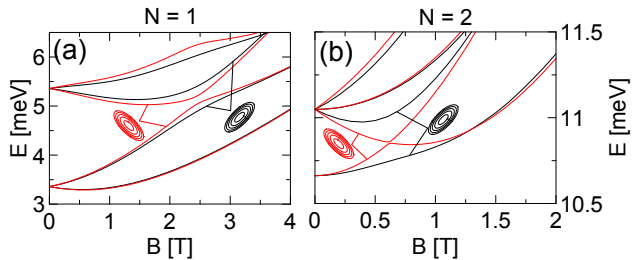}
                 \caption{One (a) and  two-electron (b) energy spectrum for elliptical dot aligned along $[110]$ (black) and $[1\overline{1}0]$ (red) for the elliptic parabolic confinement potential given by Eq. (\ref{para}). Results are obtained for $\alpha=\beta=10.8$ meV nm. The insets present the equipotential lines.}
 \label{harm}
\end{figure}

\begin{figure}[ht!]
\epsfysize=40mm
                \epsfbox[17 304 574 570] {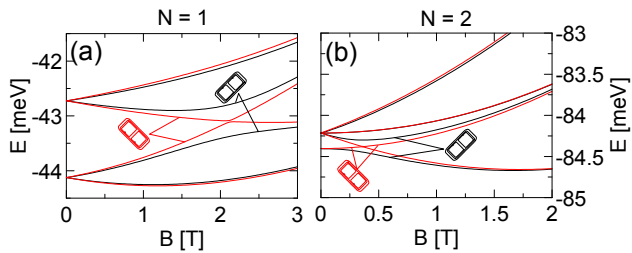}
                 \caption{Same as Fig. \ref{harm} only for the double dot potential (16).}
 \label{dd}
\end{figure}

Confinement potential of electrostatic quantum dots \cite{eqd} is generally parabolic close to its minimum,
although potential profiles closer to a quantum well can also be realized.\cite{bednareklis}
We considered the potential in form
\begin{eqnarray}
V_p(x',y')& =&\frac{m^*\omega_x^2}{2} x'^2+\frac{m^*\omega_y^2}{2} y'^2\label{para},
\end{eqnarray}
with $\hbar \omega_x=2$ meV and $\hbar \omega_y=5$ meV.
The results for $x$' identified with [110] and $[1\overline{1}0]$ directions are given in Fig. \ref{harm}. The avoided crossing between the first and second excited states of the single-electron [Fig. \ref{harm}(a)]
as well as the singlet-triplet [Fig. \ref{harm}(b)] avoided crossing  vary strongly with the orientation of the dots in consistence with the above discussion for the
confinement potential given by Eq. (\ref{prime}).

We model a double dot by introducing a barrier in the center of the quantum dot
\begin{eqnarray}
V_d(x',y')& =&V_c(x',y')\\&+& \nonumber \frac{V_{b}}{\left( 1+ \left[ \frac{x'^2}{K^2} \right]^\mu \right) \left( 1+ \left[ \frac{y'^2}{B^2} \right]^\mu \right)},
\end{eqnarray}
where $V_c$ is defined by Eq. (\ref{prime}) with $2K=40$ nm and $2L=90$ nm, $V_b=10$ meV, and $2B=10$ nm is taken for the barrier width.
For comparison the results for the single dot were presented in Fig. 3(c) for the single electron and in Fig. 6(a) for the electron pair.
Both the single- and two-electron avoided crossings that were discussed above involved mixing of even and odd spatial parity states by SO interaction. For the double dot these states correspond to bonding and antibonding orbitals, respectively. The avoided crossings observed for the double dot
are considerable thinner than in the single-dot case which is due to the introduction of the interdot barrier (in the limit of an impenetrable interdot barrier the bonding and antibonding orbitals are degenerate). In Fig. \ref{dd} we find the dependence of the width of avoided crossing on the orientation of the dot that agrees with the precedent results.

\subsection{Discussion}
The presented results indicate that the
width of SO-related avoided crossings can be designed  by specific orientation of the dot with respect to the crystal axes.
The choice of the orientation has to be done at the sample fabrication stage.
In gated quantum dots with confinement potential of the electrostatic origin \cite{eqd} the orientation of the dot can be chosen quite arbitrarily
by the shape of electrodes defined on the sample surface.
Orientation of quantum dots with structural confinement can also be intentionally controlled.
For instance, InGaAs/GaAs double quantum dots are formed on pre-patterned substrates along either [110] or $[1\overline{1}0]$ directions,\cite{lwangnjp} for which the width of the SO-related avoided crossing acquires extremal values.
In electrostatic quantum dots with a multielectrode setup,\cite{eqd} rotation of the confinement
potential should be possible to realize by voltages applied to the gates on a single sample.

We find that the orientation of the dot influences the width of avoided crossings provided that both the linear SO coupling constants are similar.
The Rashba interaction constant can be adjusted
by external electric fields,\cite{psali} to match the Dresselhaus constant in particular for observation \cite{pshe} of persistent spin helix states in quantum wells.\cite{psht}
When only a single type of SO coupling is present the orientation of the dots has no influence on the energy spectrum.
In presence of the SO coupling the orbital angular momentum is not a good quantum number even for  circular confinement potentials.
Nevertheless, both the spin components of single-electron wave function do possess a definite albeit different angular momenta.
In consequence the charge density reproduces the circular symmetry of the confinement potential.\cite{nowakring}
In potentials of circular symmetry the charge density becomes anisotropic only when both SO coupling types are present
and additionally the Zeeman effect is introduced by external magnetic field.\cite{nowakring}


The effect of the dot orientation on the energy spectrum is obtained
in external magnetic field.
For $B=0$, the dot orientation has no influence on the Kramers-degenerate energy spectrum even when both
SO  coupling types are present.  For $\alpha=\beta$ the effective magnetic field introduced by SO coupling is oriented along
the $[1\overline{1}0]$ direction independent of the orientation of the dot, nevertheless its strength is dot-orientation-dependent.\cite{nowaktd}
The electron spins precess in the effective magnetic field. Therefore, the orientation of the dot does matter for the spin-manipulation
at zero magnetic field,\cite{com,nowaktd} even though no effect on the energy spectrum is present.

\section{Summary and Conclusions}
We studied the avoided crossings opened by SO interaction in the single- and two-electron planar (001) quantum dots
as functions of the external perpendicular magnetic field.
We demonstrated that the width of these avoided crossings can be tuned within a range of two orders of magnitude
by orientation of the quantum dot with respect to the crystal directions.

The tunability is achieved provided
that  i) both Rashba and Dresselhaus interactions are present with comparable values of linear coupling constants ($\alpha\simeq \beta$)
ii) the dot is anisotropic  and iii) its larger length is comparable to
$\lambda_{SO}={\pi\hbar^2}/{2\alpha m^*}$.
The dependence of the width of avoided crossings
on the orientation of the dot results from a different strength of the Zeeman interaction
which more or less efficiently polarizes the electron spin. The spin polarization removes the spin-orbital entanglement from wave functions along
with the SO coupling effects from the energy spectrum.
Thus, the dot orientation affects simultaneously the width of avoided crossings and the effective Land{\`e} factor $g^*$.
As a general rule, the dot orientations producing large $|g^*|$ values correspond to narrow avoided crossings.

{\it Acknowledgements.}
This work was supported by the "Krakow Interdisciplinary PhD-Project in Nanoscience and Advanced Nanostructures'' operated within the Foundation
for Polish Science MPD Programme co-financed by the EU European Regional
Development Fund.  Calculations were performed in
ACK\---CY\-F\-RO\-NET\---AGH on the RackServer Zeus.


\begin{thebibliography}{00}
\bibitem{note2}
For $2K=40$ nm and $2L=90$ nm [see Fig. 3(c)], $\alpha=\beta=10.8$ meV nm the four lowest eigenvalues of Hamiltonian (1) are (in meV):
for B=0: -45.5662,-45.5662,-42.7978,-42.7978,
and for $B=5$ T -44.3195,-43.0897,-41.6678,-41.1676. When one removes the cubic Dresselhaus
and diagonal Rashba terms of Eq. (1) leaving only the linear terms of SO interaction,
one obtains for 0T:
 -45.5661,-45.5661,-42.7996,-42.7996, and
for 5T: -44.3061,-43.0753,-41.6817,-41.1612, respectively.
\bibitem{note} Note, that the results with lines and symbols of Fig. 2 are obtained for different Hamiltonians. The symbols
of Fig. 2 are obtained by diagonalization of matrix (\ref{matrix}) operator in which the magnetic field enters {\it exclusively} through the Zeeman
term. The lines are obtained for the exact Hamiltonian, in which the magnetic field modifies also the spatial operator $h$ and the SO coupling terms. Neverheless, this modification is very weak in the range of the magnetic field considered for the discussion of Fig. 2.
\end{thebibliography}
\end{document}